\documentclass[aps,prc,final,
showpacs,floatfix,amsmath,amssymb,]{revtex4}
\usepackage{graphicx}

\begin{document}

\title{Superscaling in electron-nucleus scattering and its link to CC and NC QE
neutrino-nucleus scattering}


\author{M.B. Barbaro}
  \affiliation{
Dipartimento di Fisica, 
Universit\`a di Torino and
  INFN, Sezione di Torino, 
10125 Torino, ITALY}

\author{J.E. Amaro}
  \affiliation{
Departamento de F\'{\i}sica At\'{o}mica, Molecular y Nuclear,
Universidad de Granada,
  18071 Granada, SPAIN}

\author{J.A. Caballero}
  \affiliation{
Departamento de F\'{\i}sica At\'{o}mica, Molecular y Nuclear,
Universidad de Sevilla, 41080 Sevilla, SPAIN}

\author{T.W. Donnelly}
  \affiliation{
CTP, LNS and Department of Physics, 
MIT,  Cambridge, MA 02139, USA}

\author{R. Gonz\'alez-Jim\'enez}
  \affiliation{
Departamento de F\'{\i}sica At\'{o}mica, Molecular y Nuclear,
Universidad de Sevilla, 41080 Sevilla, SPAIN}

\author{M. Ivanov}
  \affiliation{
Institute for Nuclear Research and Nuclear Energy, 
 Sofia 1784, BULGARIA}

\author{J.M. Ud\'{\i}as}
  \affiliation{
Grupo de F\'{\i}sica Nuclear, Departamento de  F\'{\i}sica At\'{o}mica, Molecular y Nuclear,
Universidad Complutense de Madrid,
  28040 Madrid, SPAIN}

\begin{abstract}
The superscaling approach (SuSA) to neutrino-nucleus scattering, 
based on the assumed universality of the scaling function for
electromagnetic and weak interactions, is reviewed.
The predictions of the SuSA model for bot CC and NC differential and 
total cross sections are presented and compared with the MiniBooNE data.
The role of scaling violations, in particular the contribution of 
meson exchange currents in the two-particle two-hole sector, is explored. 
\end{abstract}

\maketitle



Several theoretical models have been developed in recent years 
in order to describe neutrino and antineutrino quasielastic (QE) 
scattering~\cite{Caballero:2005sj,Leitner:2006ww,Amaro:2006if,Benhar:2010nx,
Martini:2010ex,Nieves:2011pp,Meucci:2011vd}.
The obvious test for a model to be used in the description of 
QE neutrino reactions is the comparison with all the available
electron scattering data at similar kinematics.
With this motivation, the Superscaling approximation (SuSA) model,
which {\it uses} the best world data for electron scattering 
to predict neutrino cross section, was proposed in Ref.~\cite{Amaro:2004bs}. 

\begin{figure}[ht]
\includegraphics[scale=0.5]{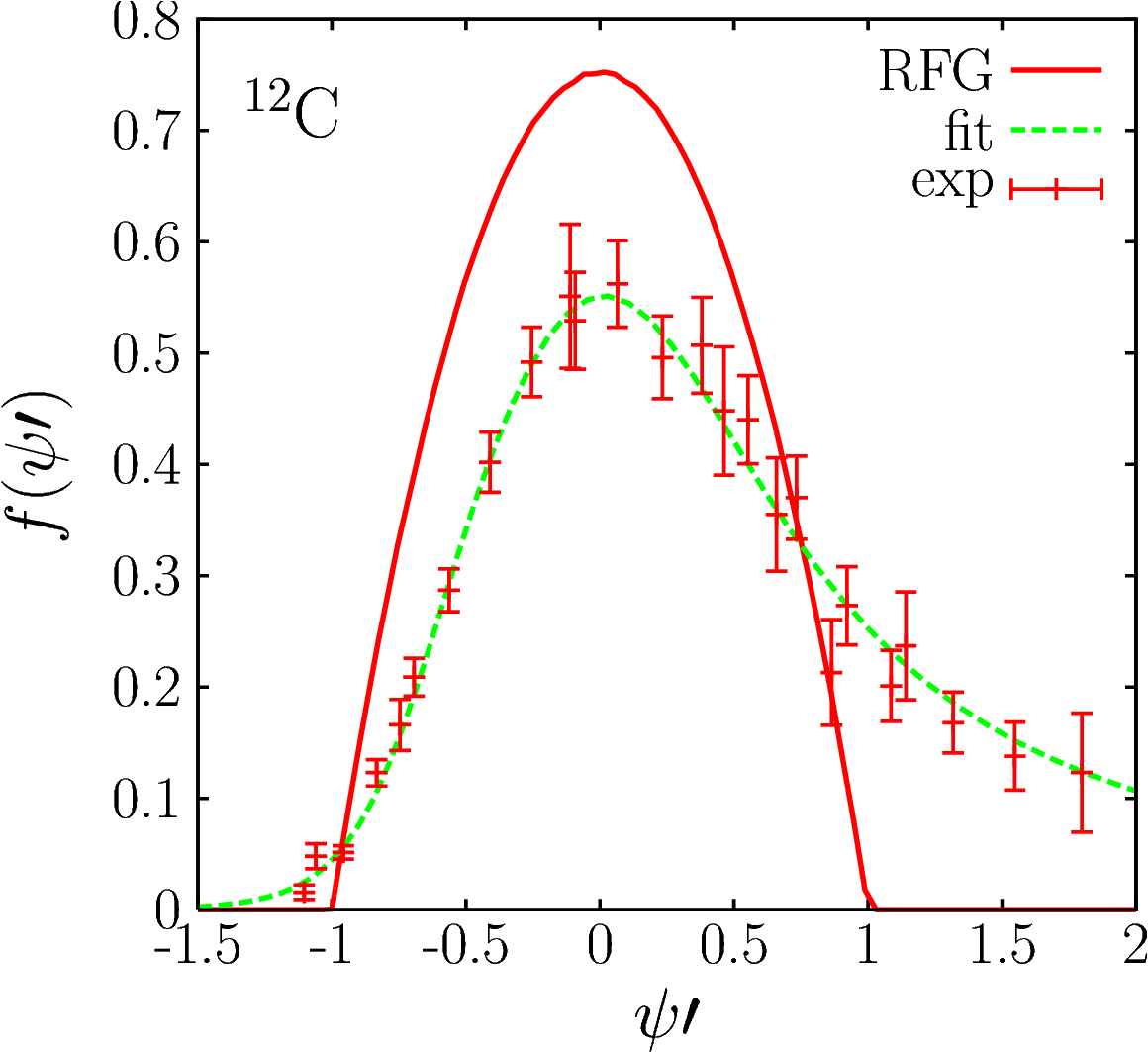}
\caption{The experimental superscaling function $f$ displayed
versus the scaling variable $\psi^\prime$
(see Ref.~\cite{Amaro:2004bs} 
for definitions) 
and compared with the RFG result.}
\label{fig:f}
\end{figure}

The model is based on the superscaling properties
of QE inclusive electron scattering data~\cite{Alberico:1988bv,
Day:1990mf,Donnelly:1998xg,Donnelly:1999sw}, namely the simultaneous 
independence of the reduced $(e,e')$ cross section upon the 
momentum transfer $q$ (first-kind scaling) and nuclear target (second-kind
scaling) when represented as a function of an appropriate scaling variable 
$\psi^\prime$.
It clearly emerges from the analysis represented in Fig.~\ref{fig:f} that the 
relativistic Fermi gas (RFG) model gives a very poor description of the data,
failing to reproduce both the size and the shape of the experimental
superscaling function.
The predictions for neutrino-nucleus cross 
sections are then simply obtained by multiplying the phenomenological
superscaling function $f$  by the corresponding elementary weak cross sections. 
The model reproduces by construction the longitudinal electron scattering 
response at all kinematics and for all nuclei.

Beyond agreeing by definition (up to scaling violations to be discussed later) 
with electron scattering data, the model has other merits.
First, it accounts for both kinematical and dynamical relativistic effects, 
which have been shown to be relevant even at moderate momentum and energy 
transfers~\cite{Amaro:1998ta,Amaro:2002mj}.
As a consequence the model can be safely used in a
wide energy range, from the MiniBooNE typical conditions ($E_\nu\simeq$ 0.7 GeV)
up to several GeV neutrino experiments as NOMAD~\cite{NOMAD}.
Second, due to second-kind scaling, it allows for a consistent treatment of different 
target nuclei, which is particularly interesting for the MINERvA 
experiment~\cite{Minerva}.
Finally, the SuSA model can be extended to non-QE kinematics, which is not the
focus of this presentation (see Refs.~\cite{Amaro:2004bs,Maieron:2009an,Ivanov:2012fm}). 

On the other hand the main limitation of the SuSA approach 
arises from the fact that it neglects scaling violations. 
These occur mainly in the transverse channel 
at energies above the QE peak and are associated to non-impulsive effects,
like inelastic scattering and meson-exchange currents (MEC).

The latter are two-body currents carried by a virtual meson
exchanged between two bound nucleons. They can excite both 1p1h and 2p2h states.
In the 1p1h sector, studies of electromagnetic $(e,e^\prime)$ process have 
shown that the MEC, when combined with the corresponding correlations, which 
are needed 
to preserve gauge invariance, give a small contribution to the QE cross section
and can be neglected in first approximation.
In the 2p2h sector, however, MEC are known to give a significant 
positive contribution to the $(e,e^\prime)$ cross section at high energy 
transfers, leading to a partial filling of the ``dip'' between the QE 
and $\Delta$-resonance peaks. 
This region is relevant for the MiniBooNE experiment, where
``QE'' events (namely with no real pions in the final state) 
can involve transferred energies far beyond the QE peak,
due to the large energy range spanned by the neutrino flux.
Such 2p2h mechanisms were indeed proposed in 
Refs.~\cite{Martini:2010ex,Nieves:2011pp} as a 
possible explanation of the ``axial mass puzzle'', namely the fact that 
the recent CCQE and NCQE MiniBooNE results~\cite{AguilarArevalo:2010zc,
AguilarArevalo:2013hm,AguilarArevalo:2010cx} turn out to be substantially 
underestimated by the RFG prediction, unless 
an unusually large {\em ad hoc} value of the axial mass 
$M_A\simeq$1.35 GeV/c$^2$  (as compared to the standard value 
$M_A\simeq$1 GeV/c$^2$) is employed in the dipole parametrization of the 
nucleon axial form factor. 
Notably also the relativistic Green's function model, 
based on a particular treatment of final state 
interaction~\cite{Meucci} through a phenomenological optical potential, 
can reasonably reproduce the MiniBooNE data, indicating that contributions 
beyond the simple IA play an 
important role in QE neutrino reactions at MiniBooNE kinematics. 

\begin{figure}[ht]
\includegraphics[width=25pc]{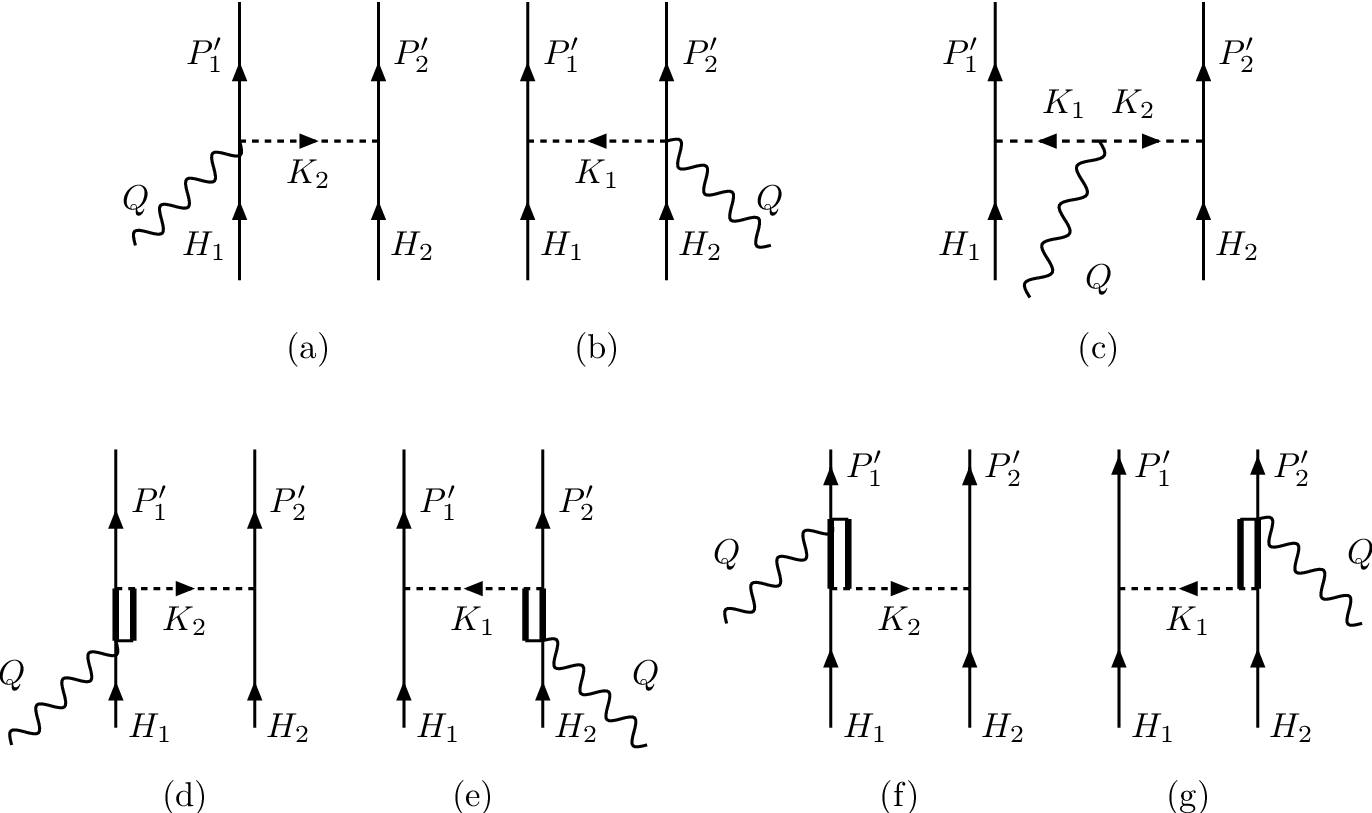}\hspace{2pc}%
\begin{minipage}{11pc}
\vspace{-5pc}%
\caption{\label{fig:mec} Two-body meson-exchange currents. 
(a) and (b): ``contact'', or ``seagull'' diagram;  
(c): ``pion-in-flight'' diagram;
(d)-(g): ``$\Delta$-MEC'' diagram. Dashed lines represent pions, thin lines
nucleon and thick lines represent the propagation of the $\Delta$-resonance.}
\end{minipage}
\end{figure}

Our approach to MEC, developed for use in electron scattering reactions (see 
Refs.~\cite{De Pace:2003xu,DePace:2004cr}), is fully relativistic and 
takes into account all the 2p2h many-body diagrams corresponding
to the two-body current in Fig.~\ref{fig:mec}. 
In order to apply the model to neutrino scattering,
we observe that in lowest order the 2p2h sector is not directly reachable 
for the axial-vector matrix elements. Hence at this order the MEC affect only
the transverse polar vector response. 
%

\begin{figure}
  \includegraphics[height=.5\textheight]{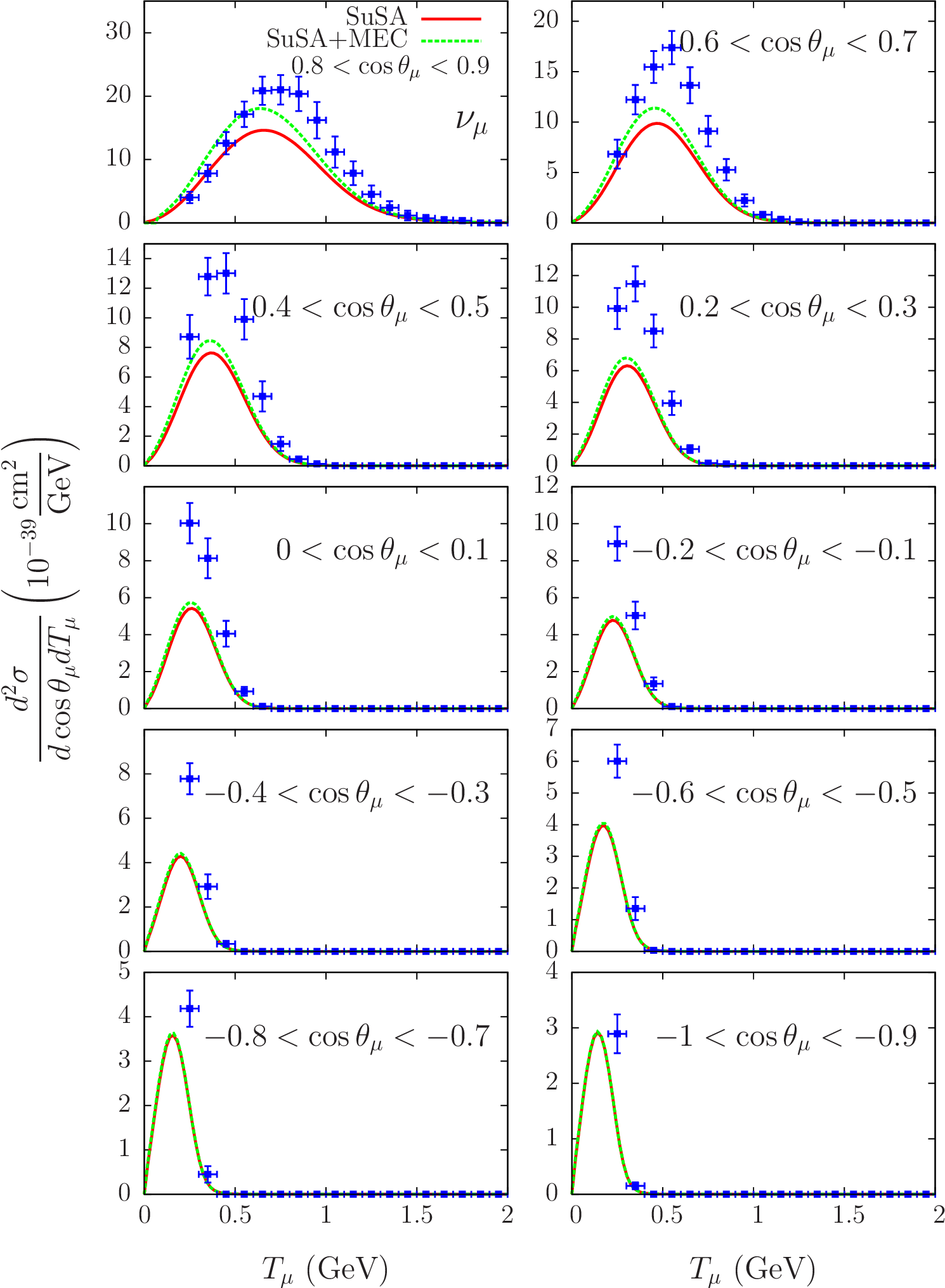}
  \includegraphics[height=.5\textheight]{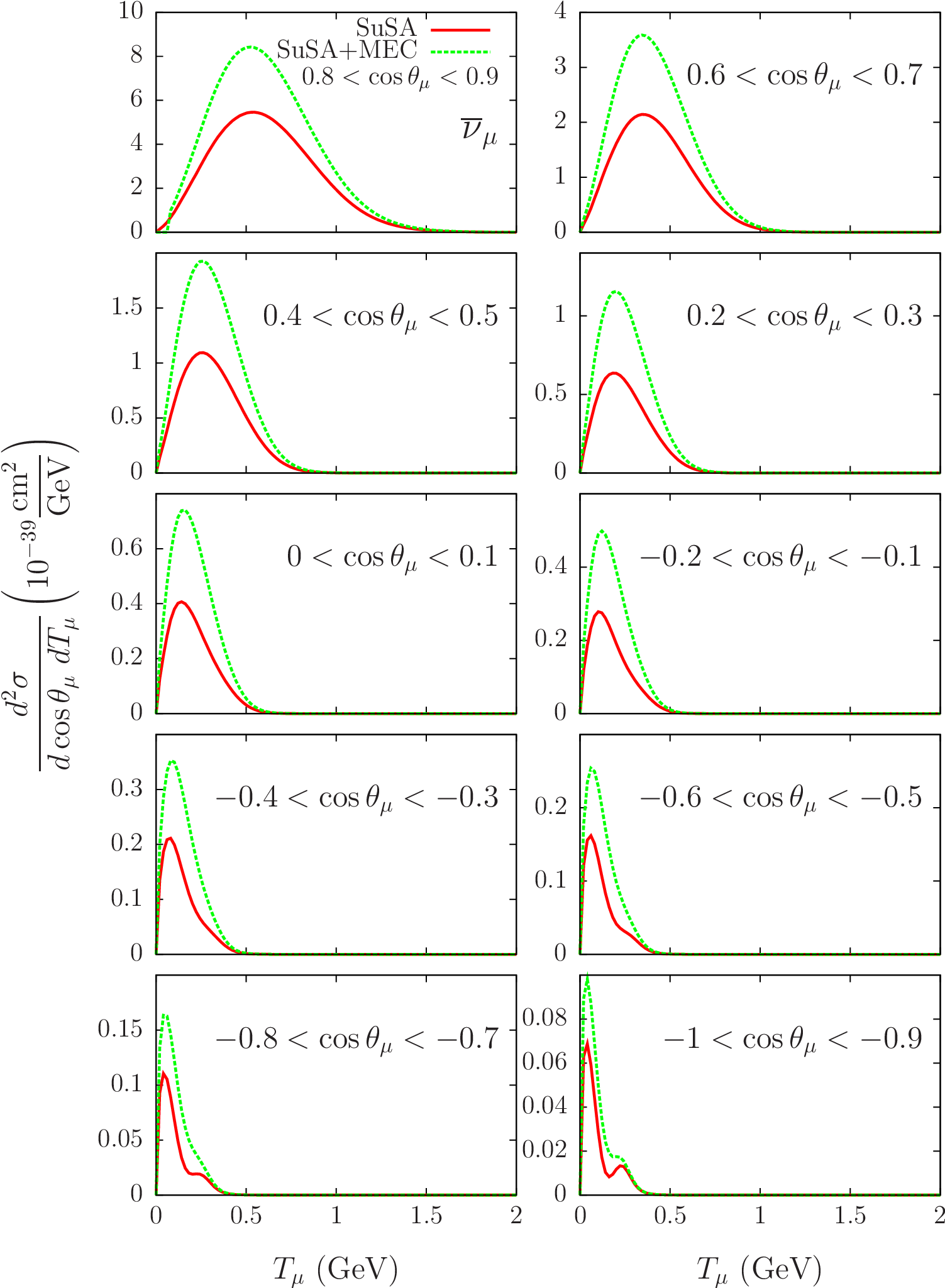}
  \caption{ Flux-averaged $\nu_\mu$-$^{12}$C (left) and
$\overline\nu_\mu$-$^{12}$C (right)  CCQE 
double differential cross sections per target nucleon 
evaluated in the SuSA model with and without inclusion of 2p2h MEC
and in the RMF model and displayed versus the muon kinetic energy $T_\mu$ 
for various bins of the muon scattering angle $\cos\theta$.
Neutrino data are from MiniBooNE~\cite{AguilarArevalo:2010zc}.} 
\label{fig:diff}
\end{figure}

The results of the SuSA model
for the double differential CCQE 
neutrino and antineutrino cross sections are shown in Figs.~\ref{fig:diff}:
they appear to fall below the neutrino data for most of the angle and energy bins.
Note that we do not compare with the most forward angles
(0.9$<\cos\theta<$1) since for such kinematics roughly 1/2 of the
cross section has been proved~\cite{Amaro:2010sd} to arise from very low 
excitation energies (<50 MeV), where the cross section is dominated by 
collective excitations and any approach based on IA is bound to fail.

The inclusion of 2p2h MEC yields larger cross sections 
and accordingly better agreement with the neutrino data, but theory still 
lies below the data at larger angles where the cross sections are smaller. 
It should be noted, however, that 
the present approach still lacks the
contributions from the correlation diagrams associated with the MEC which
are required by gauge invariance, and these might improve the agreement
with the data.

Predictions for antineutrino cross section are also shown, which can be compared
with the new antineutrino data from MiniBooNE~\cite{AguilarArevalo:2013hm}. 
Note that in our model, due to the interplay between the vector and
axial-vector responses in the total cross section, the effect of MEC is
larger than for the neutrino case.

\begin{figure}
  \includegraphics[height=.22\textheight]{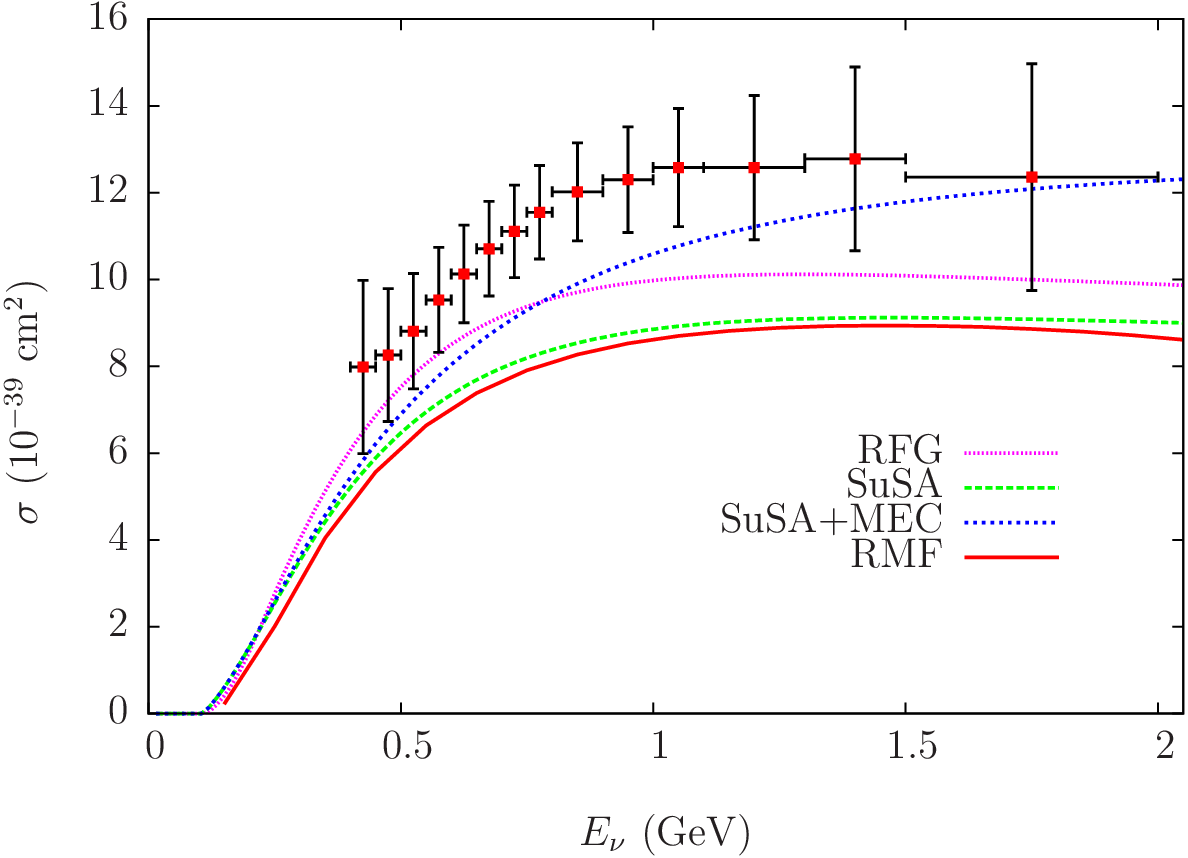}
\ \ \ \ \ \ 
  \includegraphics[height=.22\textheight]{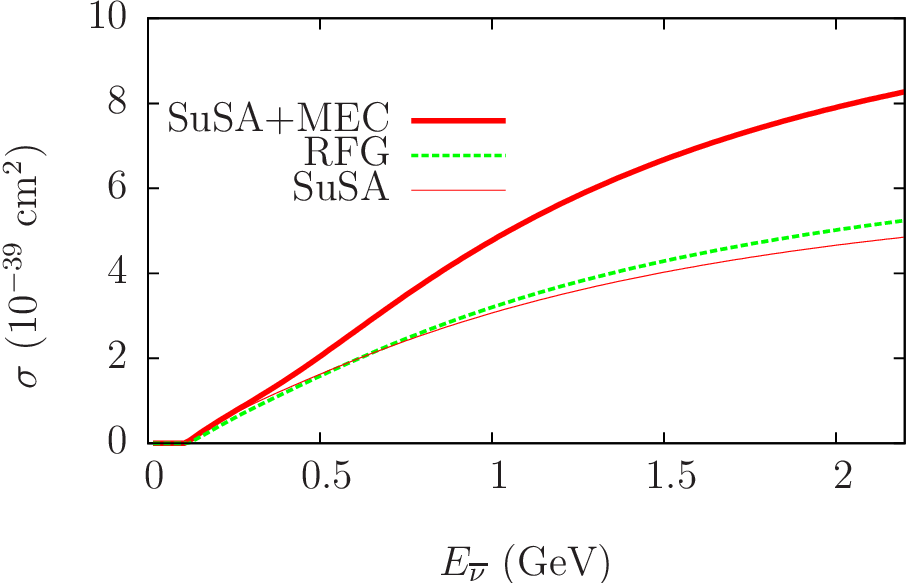}
  \caption{Flux-averaged $\nu_\mu$-$^{12}C$ (left) and
$\overline\nu_\mu$-$^{12}C$ (right) 
CCQE cross section 
integrated over the muon kinetic energy and scattering angle and displayed
versus the unfolded neutrino energy. 
Data from Ref.~\cite{AguilarArevalo:2010zc}.}
\label{fig:nutot}
\end{figure}

In Fig.~\ref{fig:nutot} we display the SuSA results for the total cross sections.
In the neutrino case we also show the results obtained in 
the relativistic mean field (RMF) model, 
which are very close to the SuSA ones. The RFM represents in fact a
microscopic justification of the phenomenological superscaling approach: 
as shown in Ref.~\cite{Caballero:2005sj}, it is able to 
reproduce the asymmetric shape of the superscaling function, a feature hardly 
described by other microscopic models.

\begin{figure}[t]
\centering
\includegraphics[height=.32\textheight,angle=270]{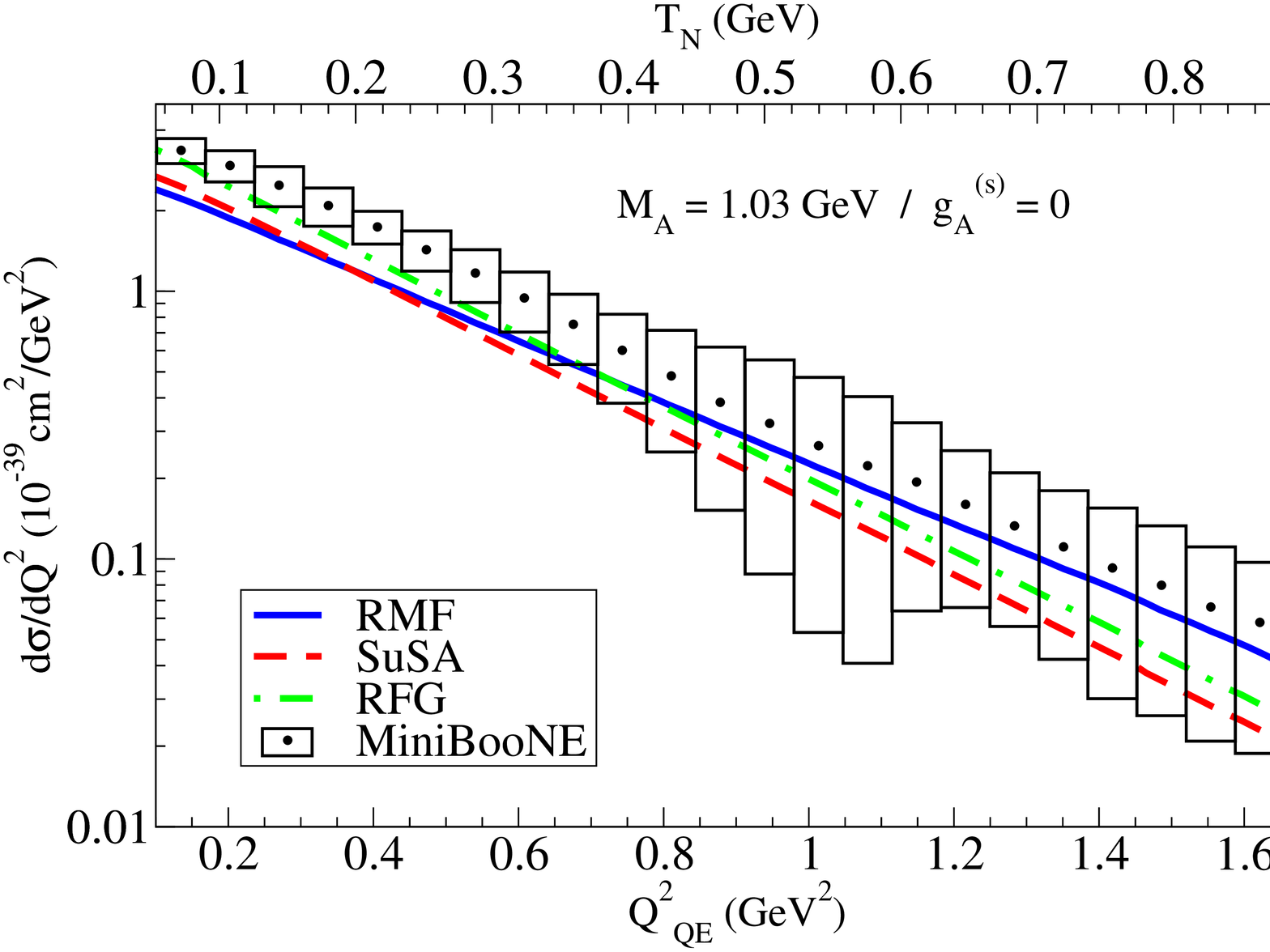}
\ \ \ \ \ \ \ \ \includegraphics[height=0.32\textheight,angle=270]{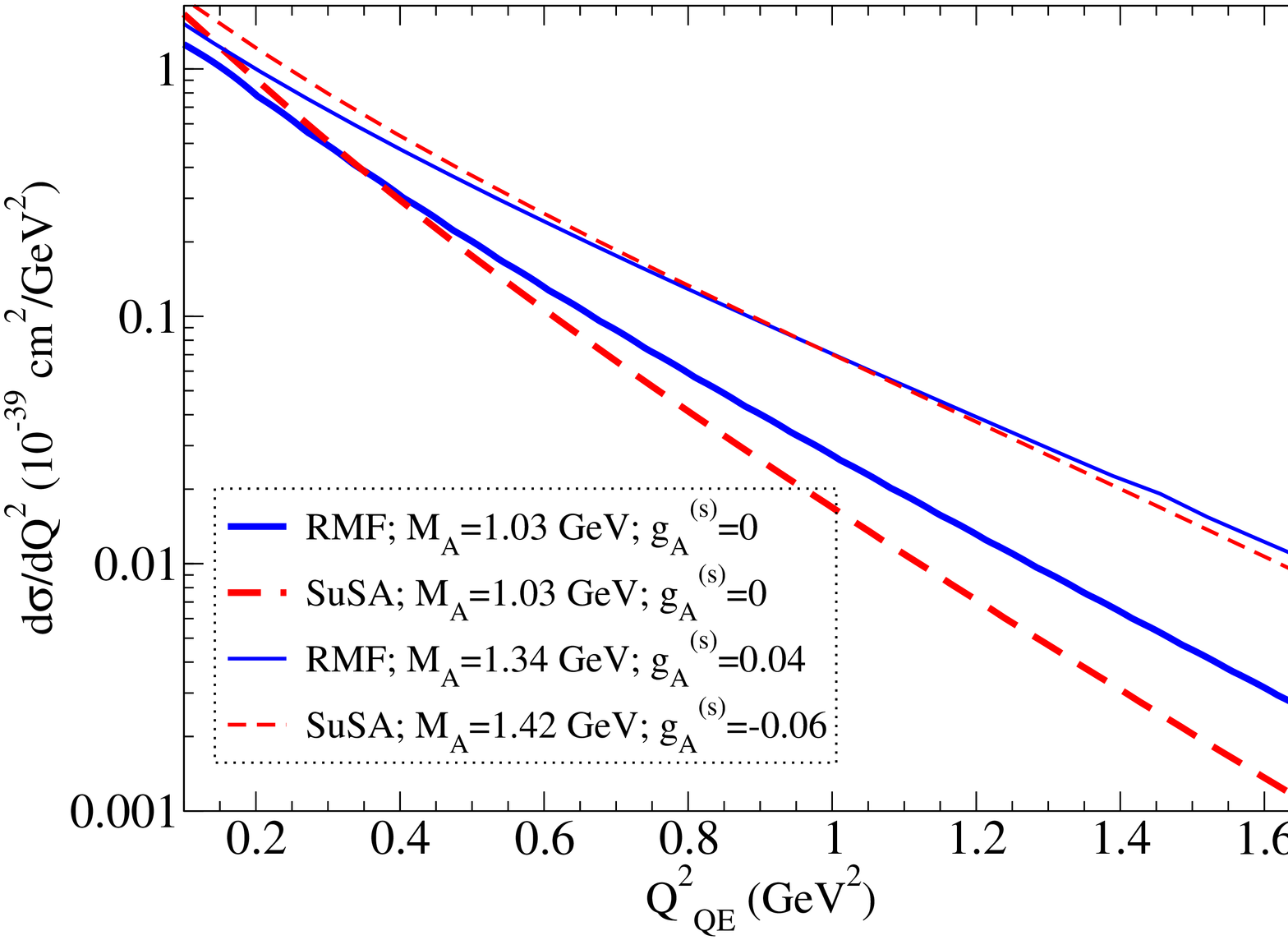}
\caption{NCQE flux-averaged neutrino (left) and antineutrino (right) cross
section off CH$_2$ compared with the
MiniBooNE data~\cite{AguilarArevalo:2010cx}.}\label{fig:CH2-flux-average-models}
\end{figure}

In Ref.~\cite{GonzalezJimenez:2012bz} 
the SuSA model has been applied to the neutral current (NC) process
with both neutrinos and antineutrinos 
and the results have been compared with the MiniBooNE data on 
CH$_2$~\cite{AguilarArevalo:2010cx}.
The results are shown in Fig.~\ref{fig:CH2-flux-average-models} as 
functions of the ``quasi-elastic'' four-momentum transfer $Q_{QE}$ defined in
\cite{AguilarArevalo:2010cx}.
We note that the SuSA cross section is smaller than the RFG one by about
20\% and the two curves have essentially the same slope in $Q^2$.
On the other hand the RMF result has a softer $Q^2$ behavior, with a smaller
slope. This is at variance with the CCQE case, for which, as shown in
Ref.~\cite{Amaro:2011qb}, SuSA and RMF cross sections are very close to each
other. This result indicates, as expected, that the NC data, for which the
outgoing nucleon is detected, are more sensitive to the different treatment
of final-state interactions than the MiniBooNE CC data, where the ejected
nucleon is not observed.

\begin{figure}[ht]\centering
         \includegraphics[width=0.3\textheight,angle=0]{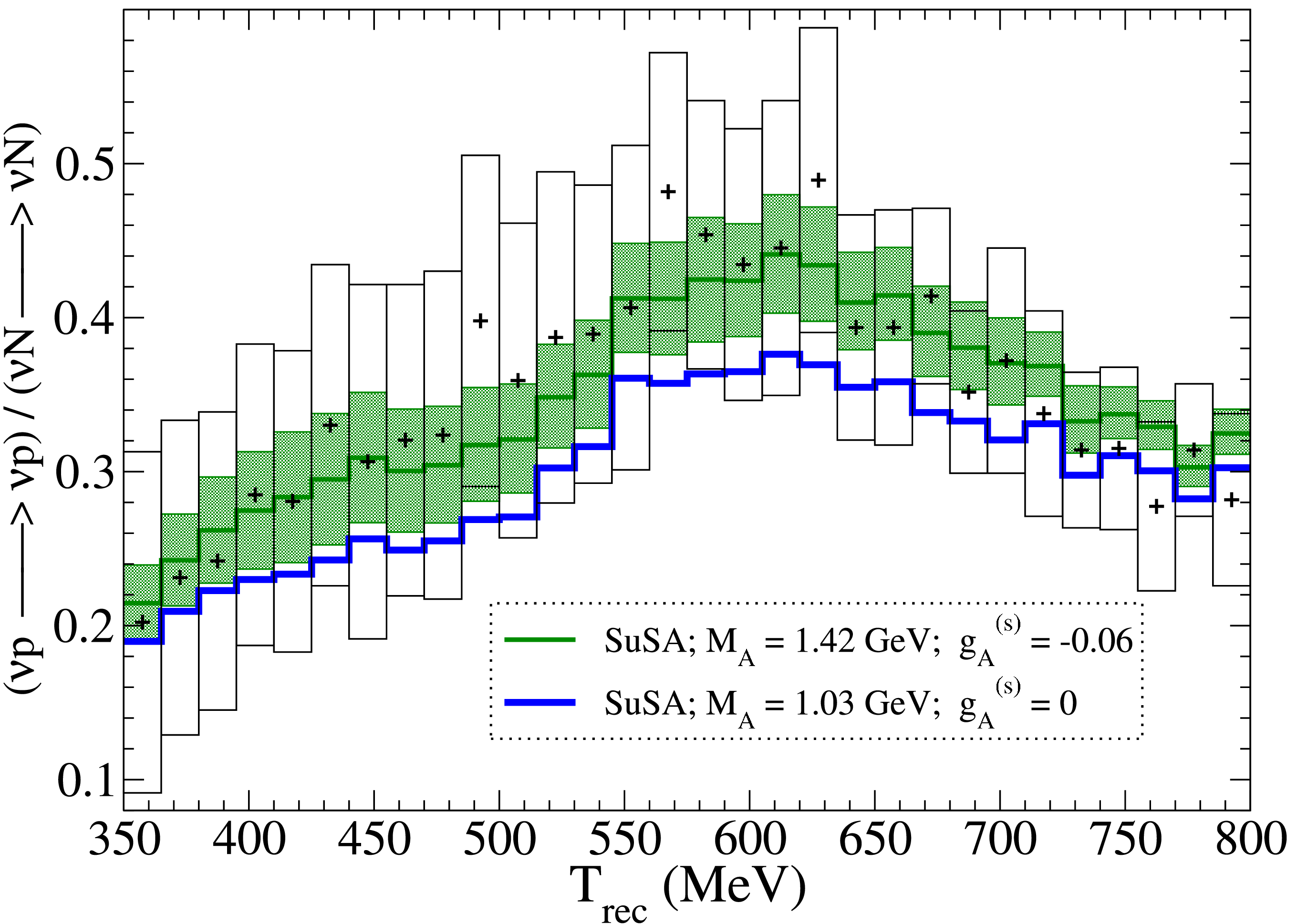}
     \caption{Ratio $(\nu p \to \nu p)/(\nu N \to \nu N)$ computed within the 
SuSA model. Shadowed
areas represent the 1-$\sigma$ region allowed for $g_A^{(s)}$. The ratio computed with the
best-$g_A^{(s)}$ is presented as well as those obtained with the standard axial mass and no strangeness.
Data from Ref.~\cite{AguilarArevalo:2010cx}.}
     \label{fig:ratio}
 \end{figure}

It is well-known (see, {\it e.g.},
\cite{Barbaro:1996vd,Alberico:1997vh}) that the $g_A^{(s)}$-dependence of the
NCQE neutrino-nucleon cross section is very mild, but the proton/neutron 
cross sections ratio is instead very sensitive to variations of the
strange axial form factor.
The MiniBooNE experiment cannot measure the $p/n$ ratio because the
$\nu n\to \nu n$ reaction cannot be isolated. However, single-proton events
have been isolated and the ratio $(\nu p \to \nu p)/(\nu N \to \nu N)$ 
has been reported in
Ref.~\cite{AguilarArevalo:2010cx} as a function of the reconstructed nucleon
kinetic energy $T_{\rm rec}$ from $350$ to $800$~MeV.
In Fig.~\ref{fig:ratio} we compare the experimental ratio with
the predictions of the SuSA model for the standard values of $M_A$ and $g_A^{(s)}$ and for a set of parameters corresponding to a  $\chi^2$ fit to the axial
strangeness parameter.
As expected, the ratio, unlike the cross section, is sensitive to axial
strangeness, and the SuSA predictions are in good agreement with the data within
the experimental error.

%


Summarizing, we have shown that the superscaling model, which by construction 
describes with excellent accuracy the QE $(e,e^\prime)$ data in the scaling
region, can be applied to the study of both CC and NC neutrino reactions. 
The model is relativistic and can therefore be used up to several GeV neutrino energies. Moreover it can be used to describe the response of different
nuclear targets to electroweak probes.

Comparison with the MiniBooNE neutrino data shows that, if the standard
value of the nucleon axial mass is used, the model underestimates the 
experimental cross sections. The inclusion of 2p2h MEC contributions,
which violate superscaling and must be added to the model, increases
both the differential and the integrated cross sections and thus 
seems to improve the agreement with the data.
However, in the present scheme,  more refined calculations taking care of 
correlation currents and MEC effects in the axial-vector channel 
should be performed before final conclusions can be drawn.
We refer the reader to the bibliography for further details and results.

\end{document}